# Thermal diffusivity degradation and point defect density in self-ion implanted tungsten

Abdallah Reza[1*], Hongbing Yu[1], Kenichiro Mizohata[2], Felix Hofmann[1†]

[1]Department of Engineering Science, University of Oxford, Parks Road, Oxford, OX1 3PJ

[2]University of Helsinki, P.O. Box 64, 00560 Helsinki, Finland

## Graphical Abstract

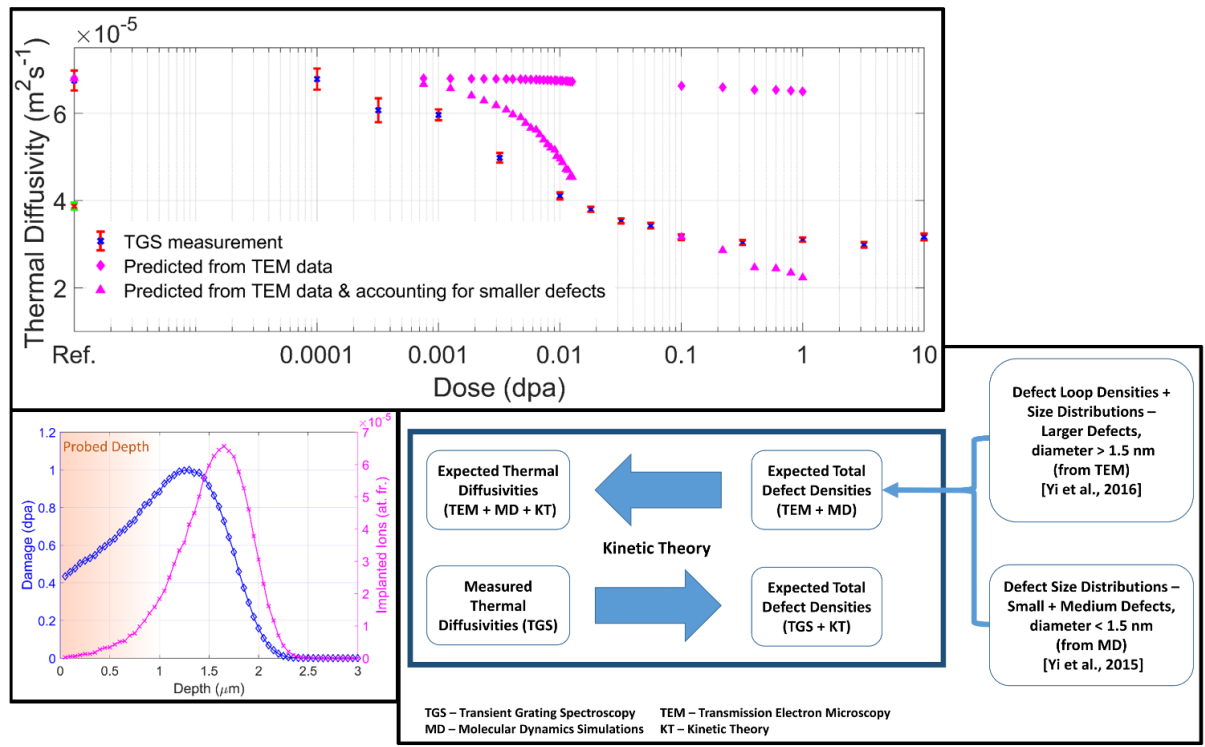

## Abstract

Using transient grating spectroscopy (TGS) we measure the thermal diffusivity of tungsten exposed to different levels of 20 MeV self-ion irradiation. Damage as low as $3.2 \times 10^{-4}$ displacements per atom (dpa) causes a measurable reduction in thermal diffusivity. Doses of 0.1 dpa and above, up to 10 dpa, give a degradation of ~55% from the pristine value at room temperature. Using a kinetic theory model, the density of irradiation-induced point defects is estimated based on the measured changes in thermal diffusivity as a function of dose. These predictions are compared with point defect and dislocation loop densities observed in transmission electron microscopy (TEM). Molecular dynamics (MD) predictions are combined with the TEM observations to estimate the density of point defects associated with defect clusters too small to be probed by TEM. When these "invisible" defects are accounted for, the total point defect density agrees well with that estimated from TGS for a range of doses spanning 3 orders of magnitude. Kinetic theory modelling is also used to estimate the thermal diffusivity degradation expected due to TEM-visible and invisible defects. Finely distributed invisible defects appear to play a much more important role in the thermal diffusivity reduction than larger TEM-visible dislocation loops. This work demonstrates the capability of TGS, in conjunction with kinetic theory models, to provide rapid, quantitative insight into defect densities and property evolution in irradiated materials.

*Corresponding author: mohamed.reza@eng.ox.ac.uk          † felix.hofmann@eng.ox.ac.uk

# 1   Introduction

A comprehensive understanding of the effect of irradiation on the properties of nuclear-relevant materials remains a substantial challenge [1–5]. Current understanding of the degradation of fission reactor material properties with irradiation has largely stemmed from empirical models derived from swathes of neutron irradiation experiments undertaken at various irradiation facilities [6,7]. Fusion reactor environments present additional challenges in the materials domain, since the radiation and heat loads will be significantly larger than in their fission counterparts [8,9]. The diverter, which removes the exhaust gases from the reactor core, experiences the largest heat and radiation fluxes [10,11]. Tungsten, due to its high melting point, high thermal diffusivity, low sputtering yield and low vapour pressure is a prime contender for the diverter armour [12]. Hence understanding the evolution of thermal diffusivity in tungsten with irradiation is of key value and is undertaken in this study.

The fusion reactor core presents a hostile environment. Apart from high heat loads, the tungsten armour will be bombarded by high energy neutrons (14.1 MeV), alpha particles, deuterium ions, as well as other species [12]. The neutrons can impart sufficient energy to displace tungsten atoms from their lattice sites. These primary knock on atoms (PKAs) will dislodge further atoms from their lattice sites, creating cascades of displacement damage [13,14]. The resulting lattice defects lead to a locally increased electron scattering rate, degrading thermal diffusivity. Neutrons also cause transmutation of tungsten into rhenium, tantalum and osmium that act as effective electron scattering sites and reduce thermal diffusivity [15]. Transmutation also causes the formation of helium, which agglomerates to form bubbles within the tungsten matrix and also decreases the thermal diffusivity [16]. Similarly the diffusion of hydrogen and its isotopes from the plasma into the tungsten matrix leads to the formation of blisters that can also cause a degradation of the thermal diffusivity [17]. A significant reduction in thermal diffusivity is of concern, as it would lead to steeper temperature gradients within components for the same heat flux. This would cause (a) higher surface temperatures and an associated increased risk of melting and (b) increase thermal stresses that are largely responsible for fatigue of armour components [10].

The complexity of the different processes associated with neutron irradiation makes it attractive to consider alternative approaches that allow a specific aspect to be singled out and analysed separately. For example alloying with transmutation products such as rhenium, tantalum and osmium has been used to study the effects of transmutation [16,18]. Self-ion irradiation is attractive for singling out displacement damage for detailed analysis [19,20].

Self-ion implantation provides a shallow damaged layer, only a few microns thick, even for implantation energies of tens of MeV. Hence bulk measurement techniques for thermal characterisation are unsuitable. Transient grating spectroscopy (TGS) [21–25] enables thermal diffusivity measurements in micron thick surface layers with high temporal (few seconds) and spatial (~ 100 μm) resolution. The effect of irradiation damage on thermal transport has been successfully investigated using ion implantation and TGS on helium implanted tungsten [16] and more recently on self-ion implanted copper [26] and silicon implanted niobium [27]. Inferring radiation damage through thermal transport measurements is novel, compared to electrical resistivity measurement techniques. Recently, thermal transport measurements using time-domain thermo-reflectance [28–30] and TGS [27] have been used to qualitatively understand the underlying irradiation damage.

In this study, we perform thermal diffusivity measurements of self-ion implanted tungsten, spanning damage levels from $10^{-4}$ displacements per atom (dpa) to 10 dpa. A kinetic theory (KT) model is used to interpret the measured thermal diffusivity in terms of the underlying point defect densities resulting from the self-ion irradiation. This data is compared to transmission electron microscopy (TEM) observations [20] and predictions from molecular dynamics (MD) simulations [31] of irradiation-induced defects in tungsten. The kinetic theory model is also used to infer the thermal diffusivity reductions expected for the defect densities reported by the TEM and MD studies. The results highlight the capability of TGS to provide quantitative information about the production of irradiation damage, as well as to enable new insights into the relative importance of different types of defects in controlling thermal diffusivity degradation. Fig. 1 shows a flowchart of the core analysis work carried out in this study.

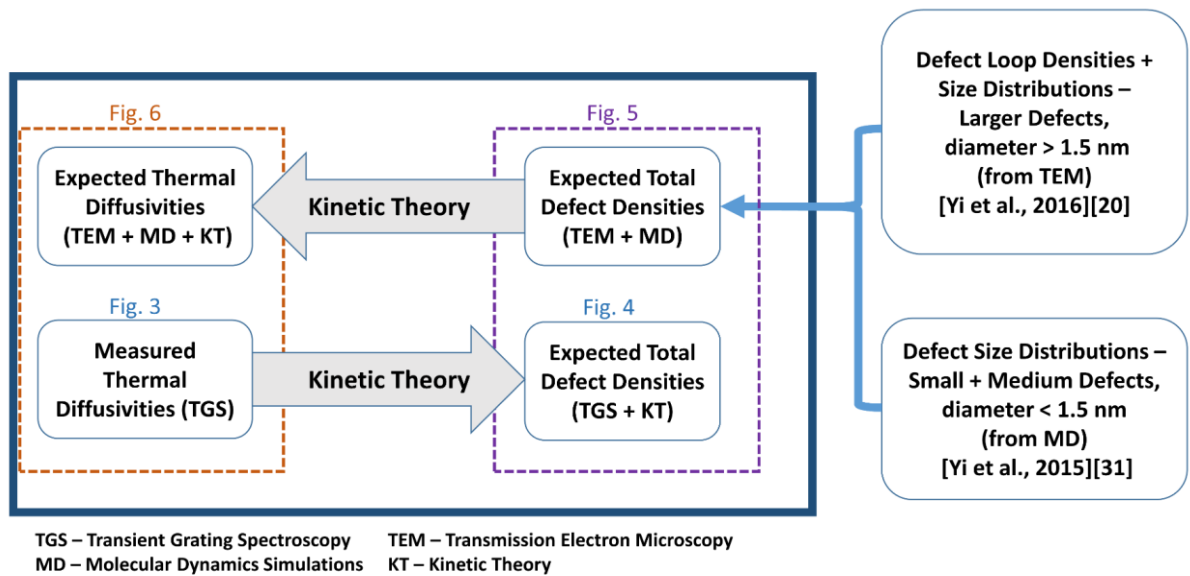

**Figure 1.** Flowchart of the core work carried out in this study, including the comparisons between the measured and expected thermal diffusivities, and TGS predicted and TEM + MD expected defect densities.

## 2  Materials and Methods
### 2.1  Sample Preparation and Implantation

Polycrystalline tungsten samples (99.97 wt% purity, procured from Plansee) 10 x 10 mm square, 1 mm thick, were annealed at 1500 °C for 24 hours in vacuum ($10^{-5}$ mbar) to allow full recrystallisation. The samples were then mechanically ground using 600 to 1200 grade abrasive papers followed by diamond slurry polishing. They were finally electrochemically polished at room temperature in a bath of 0.1 % NaOH in deionised water for 1-2 minutes at 8V to remove any polishing-induced defects and produce a high quality mirror finish. Electron back scatter diffraction (EBSD) measurements were carried out to examine the microstructure using a Merlin FEG SEM. Fig. 2(a) shows a representative EBSD map.

Ion implantations at room temperature were carried out at the Helsinki Accelerator Laboratory with 20 MeV $^{184}$W$^{5+}$ ions with a 5 MV tandem accelerator [32]. Raster scanning was used to obtain a spatially uniform implantation profile across a sample area of 15 x 15 mm$^2$, with a beam spot size of ~5 mm. A beam profilometer was used to monitor the beam current and dose. The beam current measurement was calibrated using a Faraday cup in the target chamber. The lowest dose samples ($10^{-4}$ and $3.2 \times 10^{-4}$ dpa) were implanted using beam currents of ~0.5 nA/cm$^2$, and high dose samples (3.2 and 10 dpa) were exposed at beam currents of ~90 nA/cm$^2$. These correspond to flux densities of $6.2 \times 10^8$ and $1.1 \times 10^{11}$ ions/cm$^2$/s respectively. The remaining samples were irradiated using a flux density of ~$3.1 – 5.0 \times 10^{10}$ ions/cm$^2$/s.

The Stopping and Range of Ions in Matter (SRIM) code [33] was used to estimate the fluences required for specific damage levels (Quick K-P calculation model, threshold displacement energy 68 eV for the tungsten target [34]). Injection of 20 MeV tungsten ions was modelled at normal incidence, averaging over 70022 ion calculations. While SRIM estimates the displacement damage created by ions, it does not account for any recombination of defects. Fig. 2(b) shows the predicted depth variation of displacement damage, as well as the implanted ion concentration for the 1 dpa sample. Superimposed on the plot is the depth probed by TGS – approximately 1 µm, which is well within the damaged layer.

For the implantations, 13 damage levels were considered from $10^{-4}$ to 10 dpa, where the damage level refers to the peak of the damage profile. An unimplanted sample, which had undergone the same polishing and annealing steps was used as a reference. An additional as-received sample (not annealed) was also considered. Table 1 details the damage levels considered and the corresponding implantation flux and fluences.

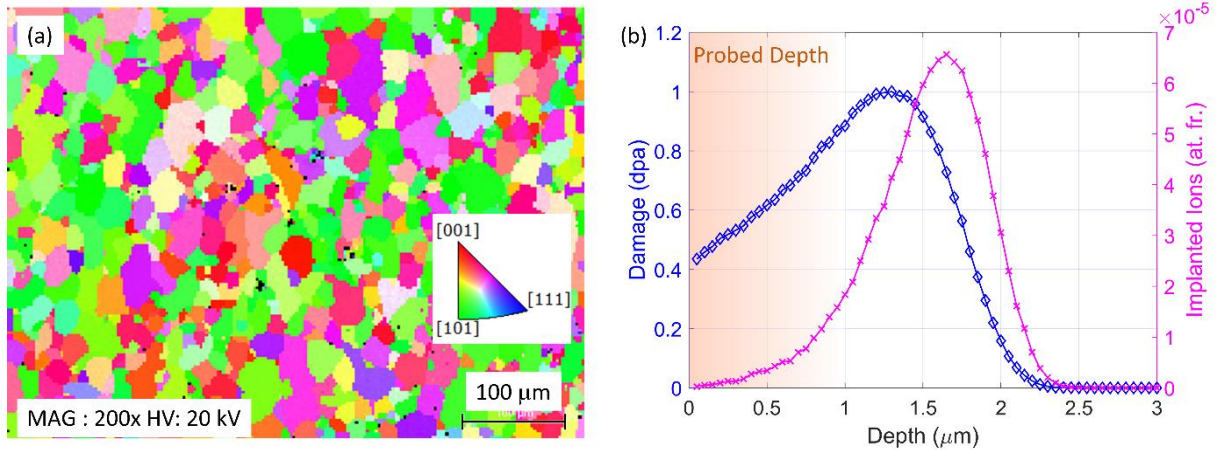

**Figure 2.** (a) Representative EBSD map of the annealed tungsten material. (b) Damage and implantation profiles predicted using SRIM for the 1 dpa sample. The TGS probing depth is shown as a shaded region.

**Table 1.** The 13 dose levels considered with their SRIM calculated fluences, incident ion fluences, beam current densities, ion flux densities and damage rates.

| Dose (dpa) | Calculated Fluence (ions/cm$^2$) | Incident Fluence (ions/cm$^2$) | Beam Current (nA/cm$^2$) | Flux Density (ions/cm$^2$/s) | Damage Rate (dpa/s) |
|---|---|---|---|---|---|
| 0.0001 | 2.53 x 10$^{10}$ | 2.7 x 10$^{10}$ | 0.5 | 6.24 x 10$^8$ | 2.3 x 10$^{-6}$ |
| 0.00032 | 8.10 x 10$^{10}$ | 8.13 x 10$^{10}$ | | | |
| 0.001 | 2.53 x 10$^{11}$ | 2.42 x 10$^{11}$ | 25-40 | 3.1-5.0 x 10$^{10}$ | 1.2 – 2.0 x 10$^{-4}$ |
| 0.0032 | 8.10 x 10$^{11}$ | 8.03 x 10$^{11}$ | | | |
| 0.01 | 2.53 x 10$^{12}$ | 2.55 x 10$^{12}$ | | | |
| 0.018 | 4.55 x 10$^{12}$ | 4.61 x 10$^{12}$ | | | |
| 0.032 | 8.10 x 10$^{12}$ | 8.20 x 10$^{12}$ | | | |
| 0.056 | 1.42 x 10$^{13}$ | 1.42 x 10$^{13}$ | | | |
| 0.1 | 2.53 x 10$^{13}$ | 2.54 x 10$^{13}$ | | | |
| 0.32 | 8.10 x 10$^{13}$ | 8.11 x 10$^{13}$ | | | |
| 1.0 | 2.53 x 10$^{14}$ | 2.53 x 10$^{14}$ | | | |
| 3.2 | 8.10 x 10$^{14}$ | 8.10 x 10$^{14}$ | 90 | 1.12 x 10$^{11}$ | 4.4 x 10$^{-4}$ |
| 10.0 | 2.53 x 10$^{15}$ | 2.53 x 10$^{15}$ | | | |

## 2.2 Thermal diffusivity measurements

Thermal diffusivity was measured using the laser-induced transient grating spectroscopy (TGS) method [16,21–25]. TGS involves excitation of the sample with two pulsed laser beams (1 kHz repetition rate, 0.5 ns pulse duration, 532 nm wavelength) that are crossed at the sample surface with a well-defined angle. Interference of the beams creates a periodic intensity pattern with selectable wavelength. Some of the light is absorbed by the sample surface, which heats up and undergoes rapid thermal expansion, creating a spatially periodic temperature and displacement grating i.e. the 'transient grating'. As heat diffuses from peaks to troughs and into the bulk, the temperature grating decays. The sudden thermal expansion of the sample also launches two counter-propagating surface acoustic waves (SAW) with the same wavelength as the excitation grating. The SAW wave speed provides information about the elastic properties of the sample [35,36]. By comparing the SAW wave speed measured from a tungsten reference sample with literature data [35–38] the TGS wavelength was calibrated and found to be $\lambda$ = 2.758 ± 0.001 µm. Tungsten is particularly suitable for use as a calibration standard since it is almost perfectly elastically isotropic [35].

The response of the sample to the excitation is probed by diffraction of a continuous wave probe beam (561 nm wavelength) from the transient temperature and surface height profiles in the sample. The diffracted probe beam is combined with a reflected reference beam (561 nm wavelength) to allow heterodyne detection. The resulting signal is recorded using a fast avalanche photo diode and oscilloscope, and analysed to obtain the thermal diffusivity and SAW wave speed data. Two probe and heterodyne beams are used in the dual-heterodyne configuration [24], which greatly increases the time resolution and signal to noise ratios. The experimental setup used here is based on the "boxcar" geometry which allows very high relative phase stability between the pump and probe beams [39]. The average probe beam power was ~22 mW (total for 2 probe beams and 2 heterodyne reference beams, chopped with a 0.25 duty cycle). The average pump (excitation) power at the sample was 1.5 mW (1.5 µJ per pulse/excitation at a 1 kHz repetition rate). The measured reflectivity was ~50%, giving the absorbed pump and probe energies by the sample as ~0.75 mW and 11 mW respectively. The excitation and probing spot sizes were ~140 µm and ~90 µm ($1/e^2$) respectively. The measurements were carried out at room temperature, and under a vacuum of ~$10^{-3}$ mbar. Sample TGS traces for implanted and unimplanted specimens are provided in supplementary figure S1. To ascertain if the TGS laser pulses affected the samples or the damage in the implanted material, the 0.32 dpa sample was measured continuously for over 5 hours. The measured thermal diffusivity remained constant over this period (see supplementary Fig. S2), indicating that the TGS measurement does not cause any modification of the sample.

Previous calculations indicate that the thermal diffusivity probed by TGS is dominated by a sample surface layer of thickness $\sim \lambda/\pi$ [21]. Thus, for the present TGS wavelength, the probed depth is slightly less than 1 µm, which is well within the thickness of the implanted layer (see Fig. 2(b)). As such the thermal diffusivity measured by TGS will be dominated by the ion-irradiated material.

# 3 Results and Discussion

## 3.1 Thermal Diffusivity degradation with dose

TGS was used to create thermal diffusivity maps of ~1-2 mm$^2$ in each sample with a step size of 150 µm, resulting in 40-50 measurement points per map. 20,000 excitations (10 measurements/point with 2,000 excitations/measurement) were sampled for each point, resulting in an uncertainty of <6% in the measured thermal diffusivity. Fitting of the experimental data was carried out in MATLAB [40] (see supplementary Fig. S3 for sample traces with fits and supplementary section 2.1 for the fitting equation). The average thermal diffusivity value from this map was then taken as the value at the corresponding damage level. Fig. 3 shows the measured thermal diffusivity values for the self-ion implanted samples for damage doses from 0.0001 dpa to 10 dpa.

Measurements of an unimplanted reference sample and of a sample that had been polished but not annealed (as rolled) are also shown in Fig. 3. The thermal diffusivity obtained for pristine, annealed tungsten is ~$6.8 \times 10^{-5}\ m^2 s^{-1}$. This agrees very well with room temperature measurements of thermal diffusivity in unimplanted tungsten from previous studies using conventional methods [15,18,41,42] and TGS [16](also shown in Fig. 3). The error bars of our measurement also fall within the range of values obtained in these studies. It is worth noting that previous thermal diffusivity measurements [41] on pure single and polycrystalline samples yielded very similar results (the samples in the present study are polycrystalline). This is expected since tungsten has a cubic lattice structure, and hence isotropic thermal diffusivity.

For comparison of the measured thermal diffusivity in ion-irradiated tungsten samples, literature data from previous tungsten irradiation experiments has also been plotted in Fig. 3: Self-ion implanted tungsten (at ambient conditions) [19]; tungsten irradiated with protons and spallation neutrons (at 115-140 °C) [43]; neutron irradiated tungsten (at 200 °C) [44]. From these studies just one is on self-ion implanted tungsten [19], and at just one dose level. The error bars observed in that study are significantly larger due to the measurement uncertainty in the 3-omega method used [19]. Fig. 3 demonstrates that TGS allows measurements with significantly lower uncertainty.

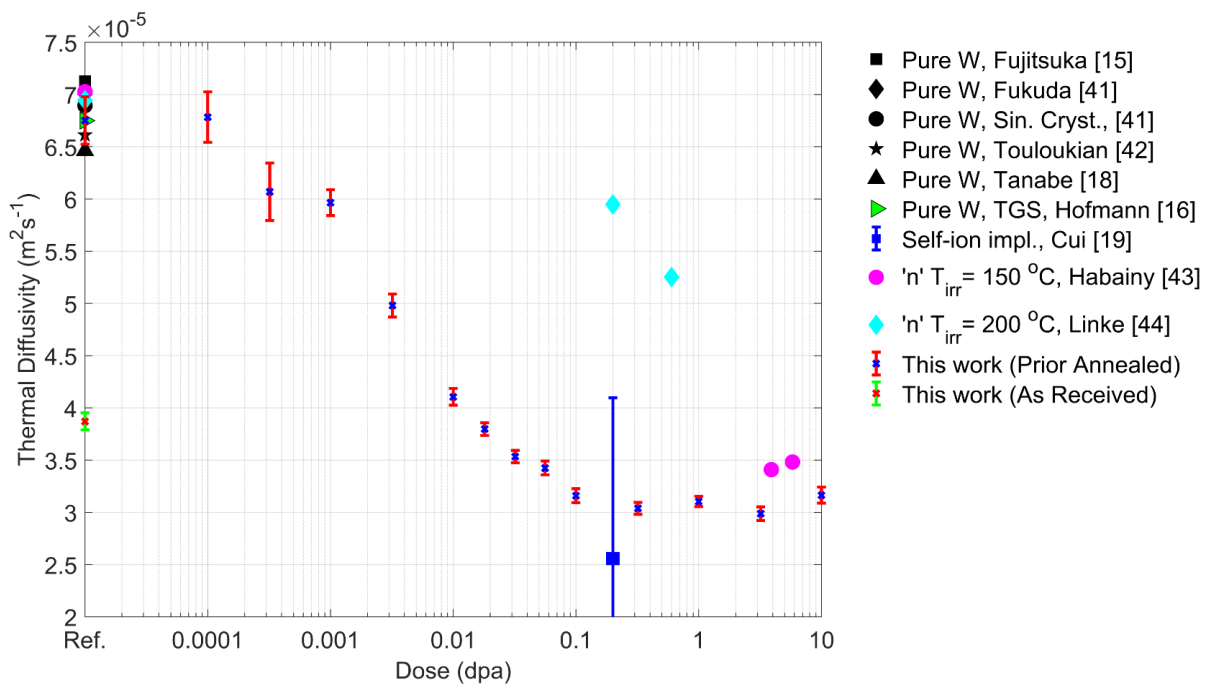

**Figure 3.** Measured thermal diffusivity values for unimplanted, as-received and self-ion implanted tungsten samples exposed to different damage doses. Literature data from previous studies on pure and irradiated tungsten is also shown.

The error bars are larger for the unimplanted, 0.0001 dpa and 0.00032 dpa samples and become progressively smaller with increasing damage level. We ascribe this to impurities in the tungsten samples. Depending on the spatial location the impurity levels will vary, giving slightly different values of thermal diffusivity. For the samples with higher damage, on the other hand, thermal diffusivity reduction due to defects is the dominant effect.

The thermal diffusivity of the implanted samples begins to drop at ~0.00032 dpa, and then reduces steadily to a value of ~$4 \times 10^{-5} \, m^2 s^{-1}$ at 0.01 dpa, a 40 % drop, after which it begins to saturate. Full saturation is achieved by 0.1 dpa, at a value of ~$3 \times 10^{-5} \, m^2 s^{-1}$, corresponding to a 55% decrease in room temperature thermal diffusivity. The thermal diffusivity then stays constant with further implantation up to 10 dpa. The dose values seen in Fig. 3 for the onset and completion of saturation agree closely with those identified for saturation of irradiation-induced defects in TEM studies [20]. This link is further explored in sections 3.2 and 3.3.

Previous data from neutron irradiated tungsten show a higher thermal diffusivity than measured here for damage levels of 0.2 and 0.6 dpa [44]. This effect may stem from the higher irradiation temperature (200 °C), which favours migration resulting in the annihilation of defects. Data from tungsten irradiated with spallation neutrons and protons [43] at a temperature of 110 – 140 °C, also shown in Fig. 3, does not exhibit this increase. It should be noted that these samples have significant helium and rhenium content. The 3.9 dpa sample is reported to have 158 ppm and the 5.8 dpa sample 245 ppm of helium. The rhenium fraction is estimated to be around 2% in both samples. Previous thermal diffusivity studies on tungsten-rhenium alloys indicate a thermal diffusivity drop of more than 30% for a 2% rhenium content [15,16]. A 10% reduction in the thermal diffusivity of tungsten has been observed for 280 ppm helium [16]. Hence it is likely that for the work in [43] the effect of the higher irradiation temperature in 'healing' some of the irradiation induced defects is countered by the helium and rhenium content accumulated in the samples. It should be noted that the rhenium content is the result of transmutation due to the neutron irradiation and the helium content is also due to the exposure. Hence, in neutron irradiated samples, there exist competing interactions that affect the thermal diffusivity. De-convolving these effects to build an effective understanding is challenging. Self-ion irradiation is attractive as it makes it possible to study the effect of displacement damage in isolation.

Fig. 3 also shows an interesting result for the as-rolled, un-annealed tungsten sample, which has a thermal diffusivity of ~$3.9 \times 10^{-5} \ m^2 s^{-1}$, 40% less than the annealed sample. Given that impurity content will be the same in both, this difference must be due to dislocations introduced by the rolling process, highlighting that crystal defects have a substantial influence on thermal transport in tungsten. The thermal diffusivity of the as-rolled tungsten is similar to that of an annealed sample exposed to a damage dose of 0.01 dpa. If un-annealed tungsten were to be used for fusion reactor armour, a reduced value of thermal diffusivity should be used for design, rather than the book value, to account for the reduced diffusivity from fabrication-induced defects.

An interesting question concerns the effect of dose rate on the irradiation induced thermal diffusivity degradation. Studies on loop and void damage in ion-implanted tungsten have shown a relatively small effect of dose rate [45]. In self-ion implanted Fe-Cr alloys, on the other hand, dose rate is seen to have a significant effect on irradiation hardening [46]. In our work, other than for the 1 x $10^{-4}$ and 3.2 x $10^{-4}$ dpa samples which had a damage rate of ~2.3 x $10^{-6}$ dpa/s, all the samples were exposed to damage rates of ~$10^{-4}$ dpa/s. If the damage rate within this range of doses were to affect the thermal diffusivity, it would be expected that the lower damage rate would cause a larger change in the material property, i.e. the thermal diffusivity, as seen in the irradiation-induced hardening study [46]. However, we do not see such an effect in the thermal diffusivity of the samples irradiated at lower dose rates. Similarly, the samples with the highest dose rates, the 3.2 and 10 dpa samples (dose rate of 4.4 x $10^{-4}$ dpa/s from table 1), would then be expected to have a higher thermal diffusivity than the 1 dpa sample. However, Fig. 3 does not show this effect. This suggests that, for the dose rates used in this study, the effect of dose rate on the thermal diffusivity is small.

## 3.2 Inferring defect populations from TGS measurements

At room temperature, electrons are the main carriers of heat in pure metals [47]. Disorder introduced into the tungsten lattice by irradiation-induced defects increases electron scattering rates, thereby reducing thermal diffusivity. In this section, the reduction in thermal diffusivity, due to ion implantation-induced damage, is used to estimate the underlying defect population.

A single tungsten atom, dislodged by a PKA, leaves behind a vacancy. The presence of a vacancy increases the scattering rate of the surrounding atoms. The dislodged atom, now called a self-interstitial atom (SIA), also has an increased scattering rate compared to when it was in the matrix. This increased electron scattering rate is responsible for the decrease in electron-mediated thermal diffusivity. A PKA streaming through the matrix creates cascades of such damage that can then interact and evolve, giving clusters and loops of interstitials and vacancies [13,48].

To infer defect populations from TGS thermal diffusivity measurements, a simple kinetic theory model for electron-mediated thermal transport is used. Here the electronic thermal conductivity is given by

$$\kappa_e = \frac{1}{3} C_e v_F^2 \tau_e \tag{1}$$

where $C_e$ is the electronic heat capacity, $v_F$ is the Fermi velocity and $\tau_e$ is the electron scattering time [49]. Considering the damaged tungsten as a dilute alloy of the pure tungsten matrix with vacancies and interstitials as 'alloying entities', the electron scattering time at a point defect (alloying atom) of type 'm' in the matrix is given by

$$\tau_{e,m} = \left(\sigma_{0,m} + \sigma_1 T + \sigma_2 T^2\right)^{-1} \tag{2}$$

where $\sigma_{0,m}$ is the impurity scattering term, $\sigma_1 T$ captures electron-phonon scattering, $\sigma_2 T^2$ is the electron-electron scattering term, and $T$ is the absolute temperature [47,50,51]. $\sigma_1$ and $\sigma_2$ are experimentally determined constants for the pure material [16], and $\sigma_{0,m}$ is the scattering rate contribution from the disorder created by the defect.

The thermal conductivity for a matrix containing various defect types, with atomic fractions $c_m$ and scattering time $\tau_{e,m}$ is then given by

$$\kappa_e = \frac{1}{3} C_e v_F^2 \left( \sum_m \frac{c_m}{\tau_{e,m}} + \left(1 - \sum_m c_m\right) \frac{1}{\tau_e} \right)^{-1} \tag{3}$$

where $\tau_e$ is the electron scattering time for the pure material [16].

Irradiation forms complex populations of defects with different sizes [13,31,52]. The exact nature of this distribution depends on irradiation conditions, temperature and impurities in the sample to mention but a few factors. This distribution is very complex and not known a-priori. Because of the large number of unknowns, including this explicitly in a scattering model is impractical.

Hence, we proceed using the simplest possible model linking defects to thermal diffusivity, concentrating on the number of point defects required to give the observed thermal diffusivity degradation. In this regard the following assumptions are made:

1. All the defects are point defects, of which there are just two types, vacancies and self-interstitials.
2. No loss of interstitials at the free surface, i.e. the vacancy and interstitial densities are equal.

3. No clustering of defects i.e. the environment of the defect is not considered. This is already an implicit assumption in the impurity scattering kinetic theory model – each atom is considered in isolation.

Rearranging eqn. 1 gives the following expression for the Frenkel pair density $c_{FP}$ in terms of the thermal diffusivity $\alpha$

$$c_{FP} = c_v = c_i = \left[ \frac{\frac{\tau_e C_e v_F^2}{3\rho C_P \alpha} - 1}{\tau_e(\sigma_i' + \sigma_v') - 2} \right] \quad (4)$$

where $\sigma_i' = 1/\tau_{e,i}$ and $\sigma_v' = 1/\tau_{e,v}$ are the adjusted scattering rates at interstitial and vacancy sites respectively. $\rho$ is the mass density and $C_P$ is the specific heat capacity. Values for $\sigma_i'$ and $\sigma_v'$ were obtained using literature data on the rate of change of electrical resistivity per interstitial and vacancy in tungsten [53]. Using the Wiedemann-Franz law, eqn. 3 can be re-written in terms of electrical resistivity $\rho_e$, Lorentz number $L$ and the defect densities as

$$\rho_e = \frac{3LT}{C_e v_F^2} \left[ \frac{c_v}{\tau_{e,v}} + \frac{c_i}{\tau_{e,i}} + \frac{(1 - c_i - c_v)}{\tau_e} \right] \quad (5)$$

By differentiating eqn. 5 with respect to the relevant defect densities $c_i$ and $c_v$, we get values for $\tau_{e,i}$ and $\tau_{e,v}$. These are then inverted to give the scattering rates for eqn. 4. Since the measurements were carried out at constant temperature, the value of $\tau_e$ remains the same throughout. Hence, it was obtained by substituting the value of the thermal diffusivity measured for the pure tungsten sample into equation 1. The remaining constants $C_e$, $v_F$, $C_P$ are obtained from literature [42,51]. Table 2 contains the numerical values and sources of the parameters used.

**Table 2.** Numerical values of parameters used and their source.

| Parameter | Value | Source |
|---|---|---|
| $\tau_e$ | 21.93 fs | Calculated |
| $\sigma_i'$ | 17.3 fs$^{-1}$ | |
| $\sigma_v'$ | 6.1 fs$^{-1}$ | |
| $C_e$ | 26208 J m$^{-3}$ K$^{-1}$ | [51] |
| $C_P$ | 132 J kg$^{-1}$ K$^{-1}$ | [42] |
| $v_F$ | 9.50 Å fs$^{-1}$ | [51] |

The Frenkel pair densities that are obtained by feeding the measured thermal diffusivities into eqn. 4 are shown in Fig. 4. The error bars show the uncertainty in the number of Frenkel defects, estimated based on the uncertainty in the thermal diffusivity measurements (see supplementary section 2.2 for details).

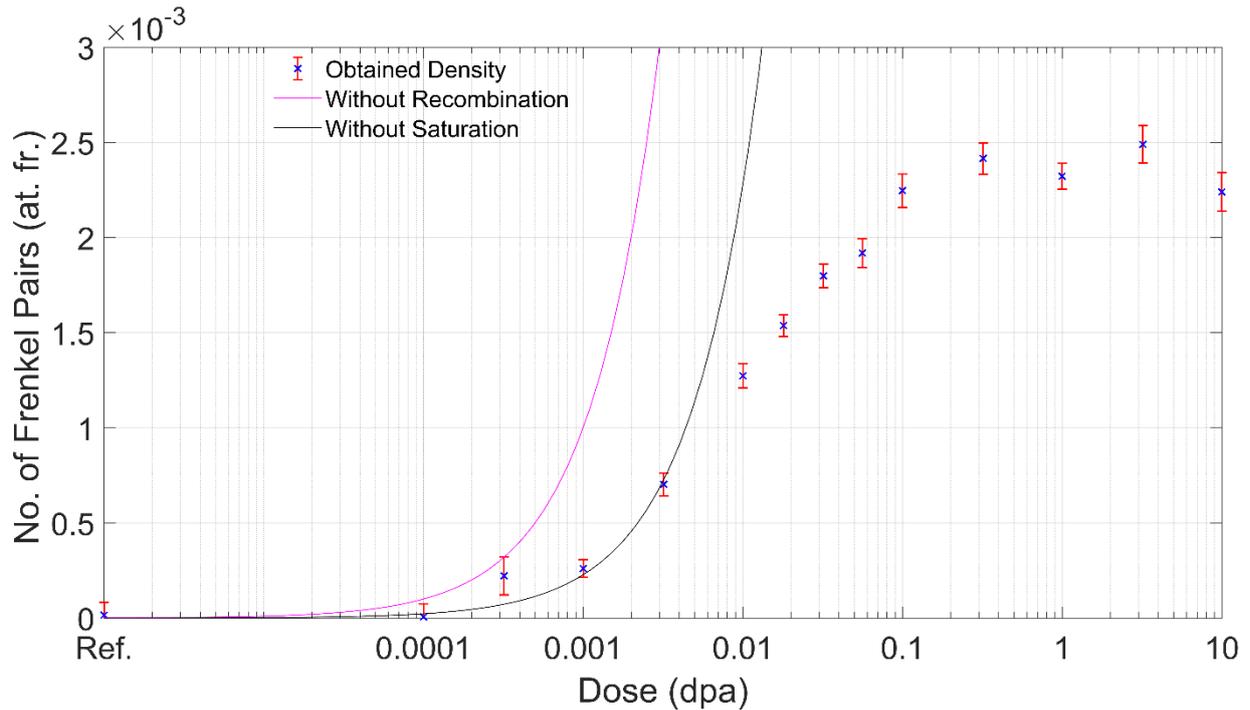

**Figure 4.** Point defect densities estimated from the measured thermal diffusivity for the unimplanted and ion-implanted tungsten samples. The magenta line denotes the defect densities expected based on SRIM calculations if there were no recombination of defects. The black curve denotes the expected densities if there were no saturation of defects. Error bars included are from the TGS measurement uncertainty only.

For doses greater than 0.00032 dpa Fig. 4 shows a noticeable increase in Frenkel defect density, which rises steadily up to an atomic fraction of defects of ~0.13% at 0.01 dpa. Above 0.1 dpa the defect density saturates at ~0.24%. This density of ~0.24% at. fr. is associated with a room temperature thermal diffusivity of ~$3 \times 10^{-5} \, m^2 s^{-1}$ (see Fig. 3).

Most of the vacancies and interstitials created by the implantation recombine. Only a small fraction of the damage created is retained. Without this recombination, the atomic fraction of Frenkel pairs in the matrix would be equal to the damage dose in displacements per atom (dpa), shown by the magenta curve in Fig. 4, i.e. all the damage created would be retained.

At lower doses a linear accumulation of damage with dose has been observed. This is the regime where there is no significant overlap of displacement damage cascades [20]. The black line in Fig. 4 shows a linear scaling of defect density with dose, fitted to damage levels up to 0.0032 dpa. It is seen to fit quite well, with a gradient of 0.23 ± 0.08 Frenkel Pairs per dpa, which is an estimation of the proportion of damage retained, also known as the NRT efficiency [54]. MD simulations of primary radiation damage in metals such as Ni, Pd, and Pt, for implantation energies of 30-200 keV are reported to give NRT efficiencies of 0.2 - 0.3 Frenkel Pairs per dpa [54]. Similar behaviour has been observed for tungsten in [55]. This NRT efficiency should translate well to higher ion energies such as the 20 MeV used in this study, due to the cascade splitting which is reported to take place above ~150 keV in tungsten [56,57]. Hence this TGS predicted defect retention estimate lies well within the bounds predicted by MD simulations. As such TGS provides a direct way of mapping out the retention of damage as a function of dose.

## 3.3 Comparison of defect densities estimated from TGS with transmission electron microscopy and molecular dynamics simulation data

It is interesting to compare the point defect densities estimated from thermal diffusivity measurements to those observed in TEM and MD simulations. TEM and MD provide information on defect cluster densities as well as damage formation and evolution. A recent detailed in-situ TEM study considered the implantation of ~150 keV tungsten ions into the tungsten matrix to mimic the effect of PKA damage [20].

A brief description of PKA damage evolution is as follows; the PKA displaces other tungsten atoms, which go on to dislodge further atoms, creating a cascade of displacement damage. Initially, this damage takes the form of a high vacancy concentration core, surrounded by interstitial atoms [13,14]. Within a few nano-seconds much of the cascade damage recombines, leaving behind a population of defects ranging from single atom (point) defects to defect clusters and dislocation loops up to several nanometres in size. These further evolve on longer time-scales through elastic interactions, as well as thermally-activated migration [31].

Previous TEM and MD studies have shown that the size distribution of this defect population obeys a power law [31,48,58]. The power law exponent is known to be between -1.6 for 400 keV ions and -1.8 for 150 keV ions [31]. However, TEM studies are not able to cover the entire range of defect sizes due to the following problem: TEM of implantation damage only permits the imaging of loops that have a diameter greater than ~1.5 nm [20,52]. Although this is larger than the ultimate spatial resolution of modern TEMs, TEM measurements are not sufficiently sensitive to probe the small contrast associated with loops smaller than 1.5 nm [59]. Since defect number density and defect size are related by a power law with a negative exponent, it is clear that whilst these "invisible" defects are small, we expect a large number density of such defects. These small defects, although invisible, can affect material properties such as thermal diffusivity at the macroscopic level.

MD simulations of self-ion damage have been successfully used to capture the creation of cascade damage and its short term evolution [31]. Specifically, for cryogenic experiments the visible defect populations predicted by MD simulations and observed by TEM are in remarkably good agreement [31]. Importantly MD simulations provide information about the whole defect microstructure, from single atom defects to large dislocation loops several nanometres in size [31,48]. Unfortunately, computational limits on the simulation time, for the time being, mean that MD simulations cannot capture the long time-scale evolution of defects, or the effect of damage accumulation during high damage dose exposures. In this section, we combine in-situ TEM results [20] with MD simulations [31] of self-ion implanted tungsten to estimate the total Frenkel pair densities for various damage levels from $10^{-3}$ dpa to 1 dpa. This data is then compared to the defect densities inferred from TGS.

TEM analysis in [20] provides the areal number density of irradiation defects (loops greater than 1.5 nm in diameter) for the dose range of 0.001 to 0.01 and 0.1 to 1 dpa and the defect loop size distribution at 0.01 dpa. The MD study in [31] contains the loop size distribution, including defects less than 1.5 nm in diameter. Combining these two gives a more complete distribution at 0.01 dpa. Considering the full defect size distribution (combined TEM & MD) the number density of Frenkel pairs (FP) at 0.01 dpa can be estimated as:

$$N_{FP\ 0.01} = \left( \phi_{0.01} \sum_{i=all\ bins} n_{0.01\ i}\ b_i\ s_i \right) \Big/ (d\ n_W)\ ,  \qquad (6)$$

where $\phi_{0.01}$ is the ion fluence in ions/m², $n_{0.01\,i}$ is the loop number density per incident ion at 0.01 dpa in a specific bin $i$, $b_i$ is the bin width and $s_i$ is the size of defects (i.e. the number of FPs in the loop) in a particular bin $i$. $d$ is the thickness of the implanted layer and $n_W$ is number of tungsten atoms per m³.

For thermal diffusivity changes, the total number of point defects within loops may not be the most relevant quantity to consider, since thermal diffusivity reduction depends on the lattice distortion associated with defects [16]. For a dislocation loop, the strain energy scales with the dislocation line length, i.e. loop circumference. As such a second scenario, where only defects on the circumference of dislocation loops are counted, might be more appropriate for comparison with TGS results. Here $N_{FPc\,0.01}$, i.e. the number of Frenkel pairs associated with the circumference of dislocation loops in the 0.01 dpa sample, is:

$$N_{FPc\,0.01} = \left(\phi_{0.01} \sum_{i=all\ bins} n_{0.01\,i}\, b_i\, c_i\right) \Big/ (d\ n_W)\ , \tag{7}$$

where $c_i$ is the number of defects around the circumference of a given loop with size corresponding to the $i^{th}$ bin. An interesting point to note is that while for large dislocations loops only a comparatively small number of point defects lie around the loop circumference, for small defect clusters the majority of constituent point defects lie on the circumference.

Unfortunately, there is no information in the literature about the variation of the defect size distribution with damage dose. Thus we assume that the defect size distribution is similar over the damage range from 0.001 to 1 dpa. The number of Frenkel pairs for a given damage level 'D' can then be obtained as follows:

$$N_{FP\,D} = \frac{m_D}{m_{0.01}} N_{FP\,0.01} \tag{8}$$

where $m_D$ and $m_{0.01}$ are the loop areal number densities in [20] for the given damage level 'D' and 0.01 dpa respectively.

$$m_{0.01} = \phi_{0.01} \sum_{i=all\ bins} n_{0.01\,i}\, b_i \tag{9}$$

Fig. 5 shows a comparison between the point defect densities estimated based on the TGS measurements and the densities obtained from TEM and MD data from [20,31].

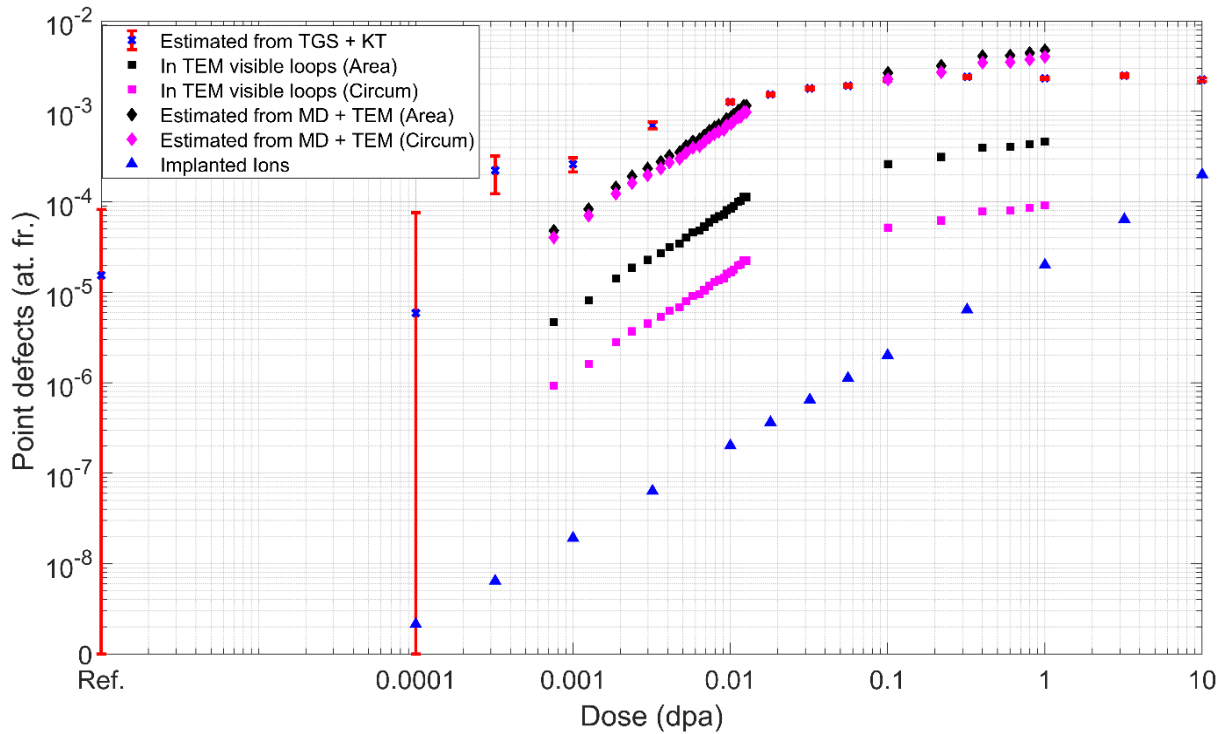

**Figure 5.** Point defect number densities estimated from TGS data with kinetic theory (KT), from TEM visible loops with the area and circumference scenarios and from combined TEM and MD data for the same scenarios. Also included in the plot are the implanted ion number densities. A loop with 10 defects is accounted for as 10 point defects in this computation. Error bars included are from the TGS measurement uncertainty only.

The densities of TEM visible defects (square markers in Fig. 5) are significantly lower than the defect densities estimated from TGS. This suggests that a significant proportion of defects is unaccounted for by TEM. Fig. 5 also shows that, for TEM visible defects, there is a substantial difference between the defect densities when all point defects within loops are considered and when only defects on the circumference of loops are counted.

Next, we consider the combined number density of defects visible in TEM, as well as point defects and small defect clusters (smaller than 1.5 nm) anticipated from MD predictions. In the low dose range (0.00032 dpa to 0.01 dpa) TEM visible defect densities vary from atomic fractions of $1 \times 10^{-6}$ at 0.001 dpa to $2 \times 10^{-5}$ at 0.01 dpa, whereas the combined TEM + MD estimate gives defect number densities of $4 \times 10^{-5}$ to $1 \times 10^{-3}$ at the same damage levels. The TGS predicted densities and combined TEM + MD predicted densities agree to within better than an order of magnitude and both rise steadily with dose. After a dose of around 0.1 dpa, the TGS predicted defect densities and the TEM + MD defect densities saturate, and match to within a factor of 2. The transition to saturation is in the same range of doses (0.01 – 0.1 dpa). Interestingly, when the smaller defects are included (TEM +MD), the circumference and area treatments differ little. The reason is that for small defects the majority of the atoms involved sit on the circumference of the defect.

The level of agreement is quite remarkable, especially considering that we are using a very simple kinetic theory model to estimate defect densities from thermal diffusivity changes. We are comparing to defect densities from TEM experiments carried out at a much lower ion implantation energy (150 keV vs 20 MeV) and using thin foil samples. The MD calculations also used a 150 keV ion energy, but only cover a comparatively short time period. The results suggest that the defects generated by ion-irradiation are in fact consistent for a given dpa level, even if quite different ion-implantation energies are used. This is in agreement with MD studies which predict that above ~150 keV in tungsten, the self-ion initiated cascades undergo cascade fragmentation and sub-cascade splitting, which implies that the higher energy PKA's will still produce cascade damage similar to that from ~150 keV implantations [56,57]. Our results also highlight that there is a large number of small point defects that are not picked up by TEM, but that significantly change material properties.

An important question concerns whether the implanted tungsten ions make a significant contribution to thermal diffusivity degradation. The implanted ion number densities, used to achieve the various damage levels as listed in Tab. 1, are superimposed on Fig. 5. Even at the highest dose, 10 dpa, the number density of injected ions is an order of magnitude lower than the defect density estimated from TGS. In the range from 0.001 to 1 dpa, where the comparison between defect densities from TGS and TEM + MD was made, the injected ion concentration is at least two orders of magnitude smaller than required to explain the measured reduction in thermal diffusivity. This suggests that the injected ions do not significantly contribute to the observed thermal diffusivity degradation. This is also affirmed by the fact that the thermal diffusivity degradation saturates at higher dose levels. The implanted ion number densities, on the other hand, increase linearly with dose. If the implanted ions did have a significant effect on thermal diffusivity such a saturation with dose would not be observed.

In addition to comparing defect densities estimated by TGS to those observed in TEM and MD simulations, eqn. 3 can also be used to estimate the anticipated thermal diffusivity based on observed or estimated point defect densities. Fig. 6 shows the predicted evolution in thermal diffusivity considering the full defect population based on TEM and MD, as well considering the TEM-visible defects only.

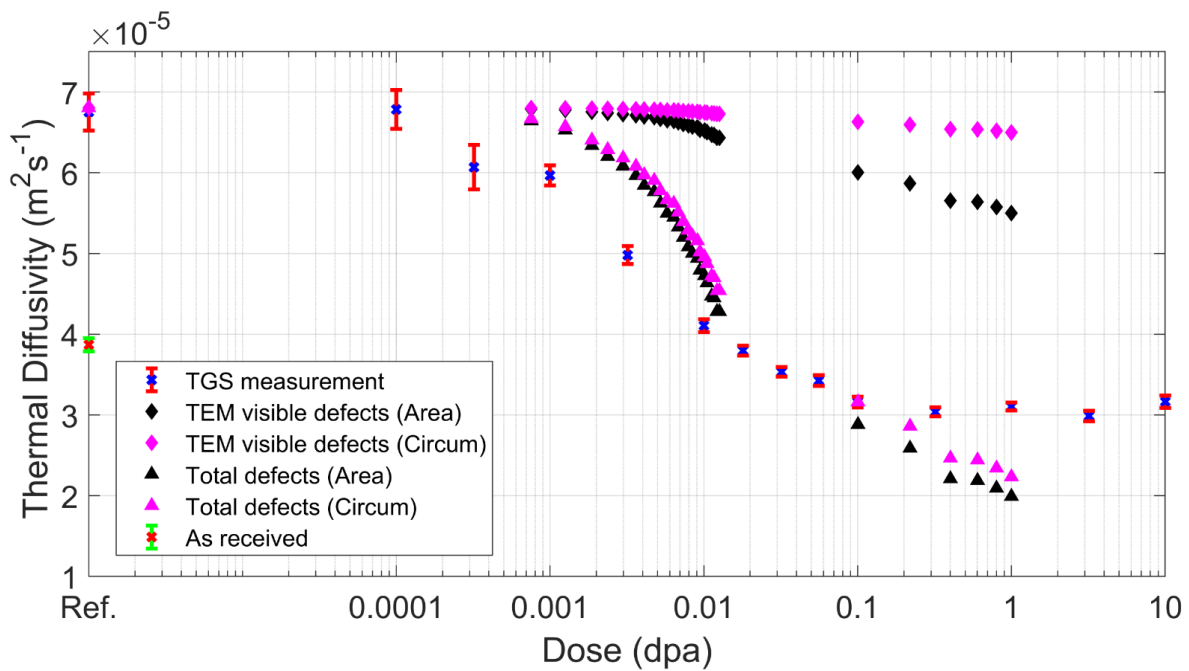

**Figure 6.** Comparison of measured thermal diffusivity using TGS, and expected thermal diffusivity values for the defect densities predicted based on TEM visible defects only and total defect densities (TEM + MD), for self-ion implanted tungsten.

Fig. 6 shows that, when only TEM visible defects are considered, the thermal diffusivity is expected to reduce to ~$5.5 \times 10^{-5}\ m^2 s^{-1}$ at 1 dpa. This is significantly higher than the actual thermal diffusivity measured by TGS of ~$3.0 \times 10^{-5}\ m^2 s^{-1}$ at 1 dpa. When the smaller defects are also accounted for, i.e. the TEM + MD scenario, the anticipated thermal diffusivity at 1 dpa is ~$2.3 \times 10^{-5}\ m^2 s^{-1}$, significantly closer to the TGS-measured value. This agreement is also seen at lower doses. As such it appears that TEM-invisible defects play a more prominent role in controlling thermal diffusivity degradation than the larger defects that are visible in TEM. This is consistent with previous studies on He-ion implanted tungsten where a substantial reduction in thermal diffusivity was observed although all defects are well below the TEM visibility limit [16].

A complication for the study of point defects and defect clusters smaller than ~1.5 nm is that they cannot be easily probed with TEM [59]. As such, most studies of irradiation damage have concentrated on the creation and evolution of defects that are visible in TEM. For example, the inability to probe small defects is the reason for the apparent incubation dose required in iron and iron alloys for the appearance of TEM-visible defects [60]. Finding techniques that easily allow the characterisation of TEM-invisible defects is challenging. Lattice swelling and electrical resistivity are sensitive to all defects but have shortcomings: Electrical resistivity methods provide only an average value over the whole specimen and are not suitable for the characterisation of few micron-thick surface layers produced by ion-irradiation [61,62]. Lattice swelling gives an integral measure of defect content [63]. However, while self-interstitials cause a lattice expansion, vacancies cause a lattice contraction. This means that the effects of interstitials and vacancies can cancel out, meaning that lattice swelling measurements will always provide a lower bound estimate of defect number density [64,65]. TGS measurements are attractive because they are sensitive to the total number of defects, since both interstitials and vacancies lead to a reduction in thermal diffusivity. In addition, TGS is sensitive even to very small defect number densities, as seen for doses as low as 0.00032 dpa in this study, and can probe few-micron-thick ion-damaged surface layers.

## 4 Conclusions

We have measured the thermal diffusivity degradation of tungsten as a result of self-ion irradiation spanning five orders of magnitude in dose from 10[-4] to 10 dpa at room temperature. Accuracy of the measurements is higher than any previous study, and the range and resolution in dose are higher than in any previous ex-situ work. Using a kinetic theory model, the measured thermal diffusivity is used to estimate the underlying point defect number density. This is directly compared to TEM observations and MD predictions of ion-implantation-induced defects in tungsten as a function of damage dose. A number of conclusions can be drawn from this work:

- Even very small irradiation doses (as low as 0.00032 dpa) decrease the thermal diffusivity of tungsten. Thermal diffusivity decreases with increasing dose, reaching a 55% reduction at 0.1 dpa. It then remains constant up to 10 dpa, the largest investigated dose.
- Defect number densities estimated from TGS are consistent with those anticipated from TEM and MD with better than order-of-magnitude agreement over three orders of magnitude in dose. This agreement is quite remarkable given the very simple kinetic theory model used and the use of different ion energy and sample geometry in ion-implantation experiments and MD simulations.
- Point defects and defect clusters smaller than ~1.5 nm play a dominant role in controlling thermal diffusivity degradation. These defects are challenging to probe since TEM is not sufficiently sensitive to detect them. Thermal diffusivity degradation is dramatically underestimated when only large defects that are visible in TEM are accounted for.

- The ability of TGS to provide quantitative defect microstructure information, in the form of point defect number densities, is demonstrated. As such TGS provides a convenient tool for probing the entire defect distribution, with excellent sensitivity even to very low defect concentrations.

It is encouraging that for tungsten, as a prototypical bcc material system, we can find consistency between material structure and physical property change. Our results highlight that it will be very important to account for irradiation-induced changes not only in mechanical properties, but also physical properties if tungsten is used as a plasma-facing armour material in future fusion reactors.

# 5 Acknowledgements

The authors would like to thank C.A. Dennett and M.P. Short from the Short Lab at MIT for their contribution towards the implementation of the TGS setup. The authors also wish to thank Sergei Dudarev and Daniel Mason from the Culham Centre for Fusion Energy, Andrea Sand from the University of Helsinki, and Xiaoou Yi from the University of Science and Technology Beijing for helpful discussions and constructive criticism. We acknowledge funding from the European Research Council (ERC) under the European Union's Horizon 2020 research and innovation programme (grant agreement No. 714697). The views and opinions expressed herein do not necessarily reflect those of the European Commission.

# Thermal diffusivity degradation and point defect density prediction in self-ion implanted tungsten - Supplementary

Abdallah Reza[1*], Hongbing Yu[1], Kenichiro Mizohata[2], Felix Hofmann[1†]

[1]Department of Engineering Science, University of Oxford, Parks Road, Oxford, OX1 3PJ

[2]University of Helsinki, P.O. Box 64, 00560 Helsinki, Finland

# 1 Supplementary Figures

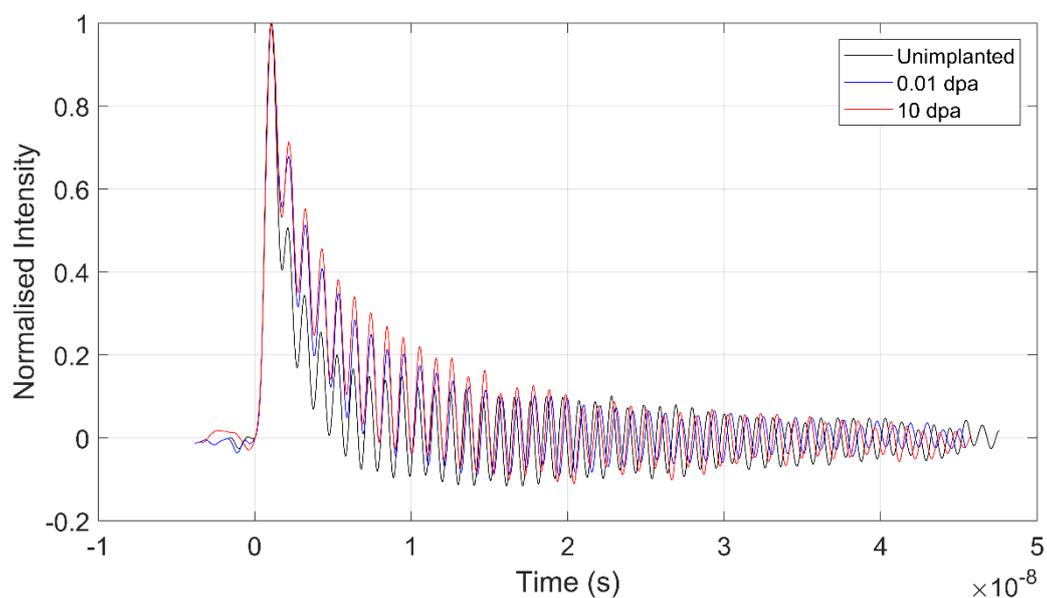

**Supplementary Figure S1.** Sample TGS traces for the unimplanted, 0.01 dpa and 10 dpa samples. The signal intensity is normalised to account for the variation due to slight changes in sample surface reflectivity. The signal decays fasters in the unimplanted sample, as the higher thermal diffusivity results in a more rapid decay of the transient temperature grating. A change in SAW frequency with implantation is also noticeable. The oscillations of the unimplanted trace are to the left of those in the the 0.01 dpa and 10 dpa traces due to the higher frequency.

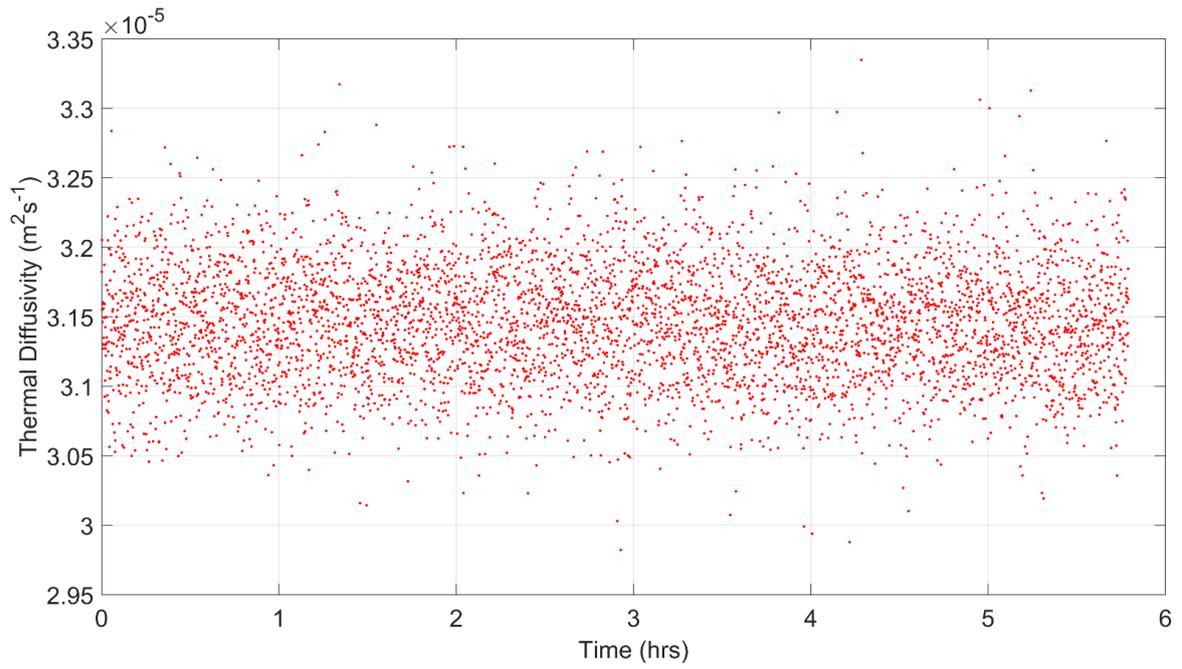

**Supplementary Figure S2.** Thermal diffusivity recorded for 5000 consecutive TGS measurements on the 0.32 dpa implanted tungsten sample. Each measurement consisted of 2000 laser pulses.

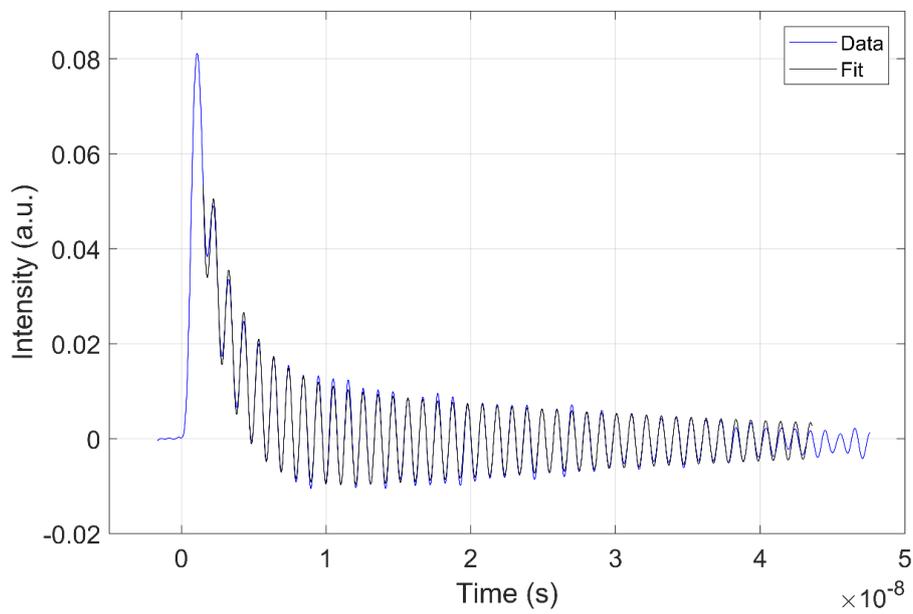

**Supplementary Figure S3.** Sample TGS trace and fit for the unimplanted tungsten sample.

## 2 Supplementary Methods

### 2.1 Fitting of Experimental Data

The TGS traces were fit to the following equation [1] using MATLAB [2].

$$I = A \operatorname{erfc}(q\sqrt{\alpha t}) + C\sin(2\pi f t + E)\exp(-t/F) + G \quad (1)$$

A, C, E, F, G, $\alpha$ and $f$ are free parameters determined by the fitting. $\alpha$ is the thermal diffusivity, $f$ is the SAW frequency, and $t$ is time.

### 2.2 Error Analysis of Defect Density Predictions

The experimental uncertainty in the measurement of the thermal diffusivity $\alpha$ creates an uncertainty in the obtained value for the Frenkel pair density. The Frenkel pair density, $c_{FP}$, is given by

$$c_{FP} = c_v = c_i = \left[\frac{\frac{\tau_e C_e v_F^2}{3\rho C_P \alpha} - 1}{\tau_e(\sigma'_i + \sigma'_v) - 2}\right] \quad (2)$$

The uncertainty in $c_{FP}$ is derived as follows:

$$\delta(c_{FP}) = \left|\frac{\partial c_{FP}}{\partial \alpha}\right| \delta(\alpha)$$

$$\delta(c_{FP}) = \left|\frac{\frac{\tau_e C_e v_F^2}{3\rho C_P}}{[\tau_e(\sigma'_i + \sigma'_v) - 2](\alpha^2)}\right| \delta(\alpha) \quad (3)$$

where $\delta(\alpha)$ is the uncertainty in the thermal diffusivity, and $\sigma'_i = 1/\tau_{e,i}$ and $\sigma'_v = 1/\tau_{e,v}$ are the adjusted scattering rates for interstitials and vacancies respectively. $\rho$ is the mass density [3], $C_P$ is the specific heat capacity [3], $C_e$ is the electronic heat capacity [4], $v_F$ is the Fermi velocity [4] and $\tau_e$ is the electron scattering time. The procedure followed in determining the electron scattering time is detailed in the main text.

# Supplementary References